\documentclass[english,reprint]{revtex4-1}
\usepackage[T1]{fontenc}
\usepackage[latin9]{inputenc}
\usepackage{amsmath}
\usepackage{stackrel}
\usepackage{amssymb}
\usepackage{graphicx}
\usepackage{subscript}
\usepackage[unicode=true,pdfusetitle,
 bookmarks=true,bookmarksnumbered=false,bookmarksopen=false,
 breaklinks=false,pdfborder={0 0 1},backref=false,colorlinks=true]
 {hyperref}
\hypersetup{
 linkcolor=blue, citecolor=blue, urlcolor=blue, pdfstartview=XYZ, plainpages=false, pdfpagelabels}
\usepackage{breakurl}

\makeatletter

\AtBeginDocument{\DeclareFontEncoding{T2A}{}{}}

\newcommand{\lyxmathsym}[1]{\ifmmode\begingroup\def\b@ld{bold}
  \text{\ifx\math@version\b@ld\bfseries\fi#1}\endgroup\else#1\fi}

\@ifundefined{textcolor}{}
{%
 \definecolor{BLACK}{gray}{0}
 \definecolor{WHITE}{gray}{1}
 \definecolor{RED}{rgb}{1,0,0}
 \definecolor{GREEN}{rgb}{0,1,0}
 \definecolor{BLUE}{rgb}{0,0,1}
 \definecolor{CYAN}{cmyk}{1,0,0,0}
 \definecolor{MAGENTA}{cmyk}{0,1,0,0}
 \definecolor{YELLOW}{cmyk}{0,0,1,0}
}

\@ifundefined{date}{}{\date{}}

\makeatother

\usepackage{babel}
\begin{document}

\title{Generalized reduction formula for Discrete Wigner functions of multiqubit
systems}

\author{K.Srinivasan}

\email{sriniphysics@gmail.com}

\selectlanguage{english}%

\author{G.Raghavan}

\email{gr@igcar.gov.in}

\selectlanguage{english}%

\affiliation{Theoretical studies section, Material Science Group, Indira Gandhi
centre for atomic research, HBNI, Kalpakkam, Tamilnadu, 603102, India.}

\date{\today}
\begin{abstract}
Density matrices and Discrete Wigner Functions are equally valid representations
of multiqubit quantum states. For density matrices, the partial trace
operation is used to obtain the quantum state of subsystems, but an
analogous prescription is not available for discrete Wigner Functions.
Further, the discrete Wigner function corresponding to a density matrix
is not unique but depends on the choice of the quantum net used for
its reconstruction. In the present work, we derive a reduction formula
for discrete Wigner functions of a general multiqubit state which
works for arbitrary quantum nets. These results would be useful for
the analysis and classification of entangled states and the study
of decoherence purely in a discrete phase space setting and also in
applications to quantum computing. 
\begin{description}
\item [{PACS~numbers}] 03.65.Ta, 03.65.Aa, {\small \par}
\end{description}
\end{abstract}
\maketitle

\section{Introduction}

Wigner distribution functions are phase space representations of continuous
variable (CV) quantum systems. These functions find widespread applications
in quantum optics. They are real valued and normalized, but unlike
genuine probability distribution functions, Wigner functions can take
negative values in some regions of the phase space, and are hence
called quasi-probability functions \citep{PhysRev.40.749,Hillery1984121}.
Classical states of light like coherent states have positive Wigner
functions \citep{HUDSON1974249} but, this is not the case for quantum
states of light such as photon added/subtracted coherent states, entangled
states and squeezed states \citep{Zavatta22102004,eltit}. In fact,
negative values of the Wigner function attest to the quantum character
of the state \citep{1464-4266-6-10-003}. Wigner functions can be
experimentally reconstructed through homodyne measurements and quantum
interference effects are quite nicely brought-out in the visual presentation
of the reconstructed state \citep{leonhardt1997measuring}. Given
the usefulness of Wigner functions of CV systems, the construction
of their finite dimensional analogs has attracted considerable attention
\citep{Hannay1980267,Bianucci2002353,Cohen1986,Feynman1987-FEYNP,PhysRevB.10.3700,Galetti1992513,0305-4470-21-13-012}.
Discrete Wigner Functions (DWFs) are particularly relevant for qubit
states used in quantum information and quantum computation studies.
DWFs find applications in stabilizer codes, quantum error correction,
quantum teleportation, study of decoherence and in the construction
of toy models in support of epistemic interpretations of the quantum
state \citep{1367-2630-14-11-113011,PhysRevA.72.012309,PhysRevA.72.042308,PhysRevA.65.062311,PhysRevA.75.032110}.
For multiqubit states, the DWF construction given by Wootters and
Gibbons et al., is particularly elegant and the present work is based
on this construction \citep{WOOTTERS19871,PhysRevA.70.062101}. DWFs
can be tomographically reconstructed through repeated measurements
using mutually unbiased basis sets (MUBS) and in the case of bipartite
system they have been directly reconstructed using Hong-Ou-Mandel
interferometers \citep{PhysRevLett.105.030406}. However, for finite
dimensional systems such as optical multiqubits, density matrices
and Stokes vectors are the most widely used representations of the
quantum state. The wider use of DWFs is inhibited by two important
limitations: (i). DWF representations of the quantum state are not
unique but, depend on the particular way of assigning the MUBS to
``lines'' in the discrete phase space, with different assignments
leading to different versions of the DWF, known as quantum nets. For
a given Hilbert space of dimension $N,$ there are $N^{N+1}$ possible
quantum nets, that is $N^{N+1}$ possible definitions of DWF. . The
state of the subsystem cannot be easily obtained as in the case of
density matrices and Stokes vectors. The problem of reduction of the
the composite state DWF to that of the sub-system has been addressed
only for two qubit systems \citep{Holmes2005}, based on the tensor
product structure of the phase space point operators. In the previous
reference, the reduction formula is given only for specific quantum
nets called the Wootters and the Aravind nets. For a $4\times4$ phase
space corresponding to a two qubit system, there are $1024$ possible
quantum nets, but phase space point operators have a product structure
only for $32$ of them and the reduction formula of Holmes et al.,
is applicable only to these cases \citep{PhysRevA.70.062101}. For
other powers of prime, the existence of the product structure of the
phase space point operator has not been investigated. In any case,
a reduction procedure for arbitrary multiqubit systems is not known
to the best of our knowledge. In the present work, we derive such
a generalized reduction formula, that does not require the existence
of such a product structure. Recently, we had addressed the problem
of carrying out spin flip operations on multiqubit DWFs \citep{1751-8121-50-8-085302}
and based on this result, we had given a formula for quantifying the
$n$-concurrence of the multiqubit systems directly from the DWF.
The relationship between the Stokes vector representation and DWF
for different choices of the quantum was exploited for this purpose
\citep{Srini20162489}. In the present work, we use some of the results
obtained in these references, to provide a general method for the
reduction of the DWF to that of its subsystems. This prescription
works for all possible quantum nets of the global system as well as
those of the subsystems. The current work is arranged as follows:
In section II we give short introduction to the DWF formalism. In
section III we discuss some earlier results which are important to
the present work, in section IV partial trace of single qubit DWF
from two qubit DWF is presented for an arbitrary quantum net as an
illustrative example of the present approach. In section V we derive
a general partial trace formula for the multiqubit systems. Sections
VI sums up the relevance of the present work as conclusions.

\section{Discrete Wigner functions}

In this section, we review the DWF construction given by Gibbons et
al., \citep{PhysRevA.70.062101}. In this approach, finite dimensional
quantum systems are represented by a discrete phase space of real
elements. A $N$ dimensional system is represented as a $N\times N$
discrete lattice of real numbers, with the points being labelled by
ordered pairs $(q,p)$ which, are elements of a finite field $\mathcal{F}_{N}$.
Since finite fields exists only for prime or prime power dimensions,
this condition imposes restrictions on the Hilbert space dimension
associated with the system. However, since our interest is in multiqubit
systems, this condition is always met. In the discrete phase space
of dimension $N\times N$, a subset of $N$-points satisfying the
equation $aq+bp=c$, for given values of $a$, $b$ and $c$ is called
a line. When the value of $a$ and $b$ are fixed, the variation of
$c$ over the finite field elements $\mathcal{F}_{N}$, generates
a set of $N$ parallel lines called a striation. In analogy with Euclidean
spaces, a set of lines are defined to be parallel if they do not share
a common point. In the $N\times N$ discrete phase space, there are
$N+1$ striations, that is $N+1$ sets of parallel lines. The point
$(0,0)$ is called the origin and any line which contains the origin
is called a ray, with each striation, containing exactly one ray.
For fixed values of field elements $x$ and $y$, if $s$ varies over
the field elements $\mathcal{F}_{N}$, the set of $N$ points $(sa,sb)=s(a,b)$
form the rays of each striation. Striations which are formed by the
fixed points $(0,1)$ and $(1,0)$ are called the vertical and horizontal
striations respectively. The remaining $N-1$ rays are formed by the
fixed points $(1,\omega)$, $(1,\omega^{2})$, $\dotsi$ ,$(1,\omega^{N-2})$.
For a given prime number $r$, there exists a finite field called
the prime field $\mathcal{F}_{r}=\left\{ 0,1,...r-1\right\} $. Finite
fields of the prime power dimensions are generated from the solutions
of the irreducible polynomial of order $r$, with prime field elements
being the coefficients of the polynomial. By defining a basis $B=\left\{ a_{1},a_{2},...,a_{n}\right\} $
for the finite field $\mathcal{F}_{N}$, every element of the finite
field $\mathcal{F}_{N}$ can be expressed as $q=\stackrel[i=1]{n}{\sum}q_{i}a_{i}$,
where the expansion coefficients $q_{i}$ are the elements of the
prime field $\mathcal{F}_{r}$. For example, let $\mathcal{F}_{2}=\left\{ 0,1\right\} $
be the prime field, then the elements of the field $\mathcal{F}_{4}$
are generated from the irreducible polynomial of order $2$, i.e.
$x^{2}+x+1=0$. If $\omega$ be one of the solution of the irreducible
polynomial, it induces the other solution $\omega^{2}=\omega+1$.
Therefore, the finite field of dimension $4$ would be given by $\mathcal{F}_{4}=\left\{ 0,1,\omega,\omega^{2}\right\} $.
For a $N$ qubit system, one may define $N^{2}$ translation operators
$T_{(q,p)}$ in discrete phase space, whose action on a line shifts
each point in the line by an amount $(q,p)$. Using the basis expansion
given above, these unitary operators are then defined as,
\begin{equation}
T_{(q,p)}=X^{q_{1}}Z^{p_{1}}\otimes\dotsi\otimes X^{q_{n}}Z^{p_{n}}\label{eq:T_Operator}
\end{equation}

where $X$ and $Z$ are the Pauli's operators and $q_{i},p_{i}\in\mathcal{F}_{r}$.
Every line in the discrete phase space is associated with a pure state,
represented by a rank one projector $Q(\lambda)$. Lines in the vertical
striations are invariant under the translation operators $T_{s(0,1)}$,
where $s$ varies over the field elements $\mathcal{F}_{N}$. Therefore,
pure states associated with the lines of the vertical striation can
be considered to be the eigenstates of the $N^{2}-1$ translation
operators $T_{s(0,1)}$. Similarly for every striation, there exists
$N^{2}-1$ translation operators which leave the lines in the striation
invariant. The state vectors associated with the lines in these striations
are simultaneous eigenstates of these $N^{2}-1$ translation operators.
However, the association of each these eigenstates to specific lines
of a striation is not unique. Each specific association is referred
to as a quantum net and leads to a different version of the DWF. For
a Hilbert space of dimension $N$, there are $N^{N+1}$ possible quantum
nets. Thus, for the quantum state represented by a density matrix
$\rho$, one may associate multiple versions of the DWF. This lack
of a one-to -one correspondence between the density matrix and the
DWF, makes the derivation of general results, independent of the quantum
net, particularly problematic. To understand the relationship between
the density matrix and the DWF, let $Q(\lambda)$ be the rank one
projector associated with the line $\lambda$. Now, the sum of the
DWF elements along this line is equal to the probability $p(\lambda)=Tr\left[Q(\lambda)\rho\right]=\underset{\alpha\in\lambda}{\sum}W_{\alpha}$.
With this association the real value taken by the DWF at each point
of the phase space is given by,
\begin{equation}
W_{\alpha}=\frac{1}{N}\left[\underset{\lambda\ni\alpha}{\sum}Tr\left(Q(\lambda)\rho\right)-1\right]\label{eq:DWF_dwf1}
\end{equation}

This can also be written as,
\begin{equation}
W_{\alpha}=\frac{1}{N}Tr(\rho A_{\alpha})\label{eq:DWF_def2}
\end{equation}

where the self-adjoint operators $A_{\alpha}$'s are the phase space
point operators, defined as,
\begin{equation}
A_{\alpha}=\underset{\lambda\ni\alpha}{\sum}Q(\lambda)-I\label{eq:PSPO}
\end{equation}

The trace product of the phase space point operators at two different
points $\alpha$ and $\beta$ is $Tr(A_{\alpha}A_{\beta})=N\delta_{\alpha\beta}$.
Therefore the set of $N^{2}$ phase space point operators can be used
as a basis for the density matrix as,
\begin{equation}
\rho=\underset{\alpha}{\sum}W_{\alpha}A_{\alpha}\label{eq:DWFtorho}
\end{equation}

The Wigner function elements are therefore the coefficient associated
with this basis expansion. One of the crucial questions that arises
in the construction of a reduction formula for the DWFs of a composite
system is whether the phase space operators have a product structure
or not. We shall see from considerations below, that even for the
simplest case of $N=4$, this is not always the case .

\subsection{Equivalence classes of quantum nets}

Two quantum nets $\mathcal{Q}$ and $\mathcal{Q}'$ are said to be
equivalent if and only if the projection operators $Q(\lambda)$ and
$Q'(\lambda)$, associated with every line $\lambda$, are related
through a unitary operator $U$. That is, there exists a unitary operator
$U$ such that for every line $\lambda$ of the phase space, $Q'(\lambda)=UQ(\lambda)U^{\dagger}$.
For example, for the $N=2$ case, if lines in the vertical and the
horizontal striations are associated with the eigenstates of the Pauli's
$\sigma_{z}$ and $\sigma_{x}$ operators, the diagonal lines would
end-up being the eigenstates of the operator $\sigma_{y}$. Now, by
assigning the states $|H\rangle$ and $|D\rangle=\frac{1}{\sqrt{2}}\left(|H\rangle+|V\rangle\right)$
to the rays of the vertical and the horizontal striations, the assignment
of the state $|R\rangle=\frac{1}{\sqrt{2}}\left(|H\rangle+i|V\rangle\right)$
or $|L\rangle=\frac{1}{\sqrt{2}}\left(|H\rangle-i|V\rangle\right)$
with the diagonal ray results in two different equivalence classes.
That is, the two quantum nets are not related through any unitary
operator $U$. Thus, there are $2$ equivalence classes in $2\times2$
phase space, and each equivalence class contains $4$ quantum nets.
Generalizing this result, a system of dimension $N$ has $N^{N-1}$
equivalence classes, where each equivalence class contains $N^{2}$
quantum nets in all. The number of equivalence classes for $N=4$
is $64$. Of these, only two of them have the special property that
the phase space point operators are tensor products of the $A_{\alpha}'s$
of the single qubit sub-systems. These operators take the form,
\begin{equation}
A_{\alpha}=A_{\alpha_{1}}^{1}\otimes\bar{A}_{\alpha_{2}}^{2}\label{eq:PSPO-EQ1}
\end{equation}

and 
\begin{equation}
A_{\alpha}=\bar{A}_{\alpha_{1}}^{1}\otimes A_{\alpha_{2}}^{2}\label{eq:PSPO-EQ2}
\end{equation}

where $\alpha=(q,p)$ and $q,p\in\mathcal{F}_{4}$. Finite field elements
$q$ and $p$ can be expressed as $q=q_{1}e_{1}+q_{2}e_{2}$ and $p=p_{1}f_{1}+p_{2}f_{2}$,
where $\left\{ e_{1},e_{2}\right\} $ and $\left\{ f_{1},f_{2}\right\} $
are the finite field basis for the horizontal and the vertical axes
and $q_{1},q_{2},p_{1},p_{2}\in\mathcal{F}_{2}$. The phase space
points of the individual systems are, $\alpha_{1}=(q_{1},p_{1})$
and $\alpha_{2}=(q_{2},p_{2})$. For these two quantum nets, the projectors
associated with each line are complex conjugates of each other. Since,
each equivalence class contains $N^{2}$ elements, there are only
$16\times2=32$ quantum nets having this special property. With this
background, we are now in a position to address the problem of obtaining
a reduction formula for the DWF.

\section{Background related to the present work}

Before discussing the reduction formula for general multiqubit systems,
we now present some results reported in our earlier papers that are
relevant to the present work. We shall provide here a summary of results
obtained for two qubit systems reported in the literature.

\subsection{Spin flipped DWF of the multiqubit systems}

In the present work, we use results from our earlier paper on performing
the spin flip operation on multiqubit DWFs \citep{1751-8121-50-8-085302}
and quantifying entanglement in such systems by exploiting the relationship
between DWFs and generalized Stokes vectors. The spin flip operation
for a multiqubit density matrix $\rho$ is defined as $\tilde{\rho}=\sigma_{y}^{\otimes n}\rho^{*}\sigma_{y}^{\otimes n}$,
where $*$ denotes the complex conjugation in the computational basis
and $\sigma_{y}$ the Pauli matrix. Let $W$ be the DWF of the multiqubit
system represented as a column vector, $W^{(*)}$ and $\tilde{W}$
are the DWFs associated with $\rho^{*}$ and $\tilde{\rho}$ respectively.
The spin flip operation can now be performed in two steps. In a first
step, the complex conjugation can be performed as $W^{(*)}=FW$, where
$F$ is a Hadamard matrix. We have shown that the Hadamard matrix
$F$ is independent of the quantum net. As a next step, the $\sigma_{y}^{\otimes n}$
operators can be considered as unitary translation operations $T_{\beta}$,
which shift each element of the discrete phase space by an amount
$\beta$. Their action on $F$ merely interchanges the rows, resulting
in another Hadamard matrix $G$. Therefore, the spin flip operation
on a multiqubit DWF can be performed by $\tilde{W}=GW$.

\subsection{Relationship between Stokes vector and DWFs of multiqubits}

In a recent work, we have given a transformation formula relating
Stokes vectors and the DWF of a given multiqubit system. We have shown
that the Stokes vectors and the DWFs of the multiqubit systems are
related through a Hadamard transformation 
\begin{equation}
S=HW\label{eq:SHW}
\end{equation}

where $S$ is the Stokes vector and $W$ the DWF (arranged as a column
vector) of the given multiqubit system and $H$ is a $N\times N$
Hadamard matrix which depends on the choice of the quantum net. For
a multiqubit systems there are $N^{N+1}$possible quantum nets, and
for each quantum net there exists unique Hadamard matrix. Let $\mathcal{\mathbb{S}}_{n}^{H}$
be the set of all Hadamard matrices for the $n$-qubit system. The
inverse of these Hadamard matrices takes the Stokes vector $S$ to
the corresponding DWF,

\begin{equation}
W=H^{-1}S\label{eq:WHS}
\end{equation}

Thus, the problem of finding the reduction formula for $W$ reduces
to that of extracting the sub-system Stokes vector and thereafter
applying the inverse of a appropriate Hadamard matrix to it.

\subsection{Reduction formula for the two qubit DWF when point operators have
a product structure}

M.Holmes et al., have given a method of performing the partial trace
operation for two qubit systems \citep{Holmes2005}. Their result
is based on the product structure of the phase space point operators
given in the work by Gibbons et al . As mentioned in section II-A,
for a two qubit systems, $A_{\alpha}$'s have a product structure
only for $32$ quantum nets. Consider the two qubit DWF defined in
the quantum net, for which the phase space point operator given by
$A_{\alpha}=A_{\alpha_{1}}^{1}\otimes\bar{A}_{\alpha_{2}}^{2}$. If
this definition is used for the reconstruction of the density matrix
given in Eq (\ref{eq:DWFtorho}), it is easy to show that the density
matrices of the subsystems $1$ and $2$ are,
\begin{equation}
\rho_{A}=\underset{\alpha_{1}}{\sum}\underset{\alpha_{2}}{\sum}W_{\alpha_{1},\alpha_{2}}A_{\alpha_{1}}\label{eq:rhoA1}
\end{equation}

and 

\begin{equation}
\rho_{B}=\underset{\alpha_{1}}{\sum}\underset{\alpha_{2}}{\sum}W_{\alpha_{1},\alpha_{2}}\bar{A}_{\alpha_{2}}\label{eq:rhoB1}
\end{equation}

respectively. It is clear from Eq (\ref{eq:rhoA1}) and Eq (\ref{eq:rhoB1})
that, the phase space point operators of the subsystems are complex
conjugates of each other. That is, the DWF of the subsystems $1$
and $2$ are defined in different quantum nets. The DWF of the first
subsystem can be calculated from Eq (\ref{eq:DWF_def2}) and Eq (\ref{eq:rhoA1})
as,
\begin{equation}
W_{\beta}^{A}=\underset{\alpha_{2}}{\sum}W_{\beta,\alpha_{2}}\label{eq:WA1}
\end{equation}

Since, the DWF of the subsystem-$2$ is defined on a quantum net where
the projection operators associated with each line have been complex
conjugated, it is necessary to perform a spin flip operation along
the $y$ direction on the DWF of subsystem $B$ i.e. $W_{\beta}^{B}$
to obtain the correct state. This can be achieved by applying the
Hadamard matrix $F$ defined in Section-IIIA by, $W^{B}=FW^{B^{(*)}}$.
Alternatively, we shall now show that this result can be achieved
in the following manner: the phase space point operator $A_{\beta}$
can be used in the place of $\bar{A}_{\beta}$ in the equation $W_{\beta}^{B}=\frac{1}{2}Tr\left(\rho_{B}\bar{A}_{\beta}\right)$
to obtain the proper DWF. This is obvious from the fact $\rho_{B}=\underset{\alpha_{2}}{\sum}W_{\alpha_{2}}^{B^{(*)}}\bar{A}_{\alpha_{2}}=\underset{\alpha_{2}}{\sum}W_{\alpha_{2}}A_{\alpha_{2}}$,
where $W_{\alpha_{2}}^{B^{(*)}}$ is the DWF associated with $\rho_{B}^{*}$.
Hence, the DWF of the subsystem-$2$ takes the form,

\begin{align}
W_{\beta}^{B} & =\frac{1}{2}Tr\left(\rho_{B}A_{\beta}\right)\\
W_{\beta}^{B} & =\frac{1}{2}Tr\left[\left(\underset{\alpha_{1}}{\sum}\underset{\alpha_{2}}{\sum}W_{\alpha_{1},\alpha_{2}}\bar{A}_{\alpha_{2}}\right)A_{\beta}\right]\\
W_{\beta}^{B} & =\frac{1}{2}\underset{\alpha_{1}}{\sum}\underset{\alpha_{2}}{\sum}W_{\alpha_{1},\alpha_{2}}Tr\left(\bar{A}_{\alpha_{2}}A_{\beta}\right)\label{eq:WB}
\end{align}

where the trace product $Tr\left(\bar{A}_{\alpha}A_{\beta}\right)$
for two different point $\alpha=\left(q_{\alpha},p_{\alpha}\right)$
and $\beta=\left(q_{\beta},p_{\beta}\right)$ is given by $Tr\left(\bar{A}_{\alpha}A_{\beta}\right)=(-1){}^{(q_{\alpha}\oplus q_{\beta})(p_{\alpha}\oplus p_{\beta})}$,
where $\oplus$-is addition modulo-$2$. Therefore, Eq (\ref{eq:WB})
can be written as,
\begin{equation}
W_{\beta}^{B}=\frac{1}{2}\underset{\alpha_{1}}{\sum}\underset{\alpha_{2}}{\sum}\left(-1\right)^{(q_{\alpha_{2}}\oplus q_{\beta})(p_{\alpha_{2}}\oplus p_{\beta})}W_{\alpha_{1},\alpha_{2}}\label{eq:WBM}
\end{equation}

From Eq (\ref{eq:WA1}) and Eq (\ref{eq:WBM}), we can calculate the
DWF of the subsystems for a given two qubit DWF. For the other equivalence
class, the complex conjugation operation needs to be performed on
the first subsystem rather than the second, with the DWF of the second
being defined on the chosen net. In this context, as Gibbons et al.,
have pointed out that the existence of the tensor product structure
is itself not established for other powers of prime. In the present
work, we provide a reduction formula for multiqubit DWFs where such
a product structure is not required. To the best of our knowledge,
such a general result is not available in the literature.

\section{A general reduction formula for the two qubit DWF for arbitrary quantum
nets}

As shown in the earlier section the approach by Holmes et al., is
restricted to only 32 of the possible 1024 quantum nets. In this section,
we derive a general result valid for all quantum nets of the global
as well as the subsystems. Let $\rho_{AB}$ be the density matrix
of the two qubit system, $\rho_{A}$ and $\rho_{B}$ be those of its
subsystems. In the density matrix representation, the subsystem can
be obtained by taking a partial trace on $\rho_{AB}$ i.e. $\rho_{A}=Tr_{B}(\rho_{AB})$
and $\rho_{B}=Tr_{A}(\rho_{AB})$. Now, to derive a formula for obtaining
the DWF of the single qubit subsystem from that of the two qubit DWF,
we need to specify the quantum net of both. Hence, the transformation
formula must be general enough to accommodate this requirement. 

Let $M$ be the observable acting on the subsystem $A$ of the general
system $\rho_{AB}$. This can be mathematically represented as $(M\otimes I)\rho_{AB}$.
The expectation value of the observable $M$ only on the subsystem
$\rho_{A}$ and the expectation value of the operator $M\otimes I$
on the global system $\rho_{AB}$ are one and the same, that is,
\begin{equation}
<M>_{\rho_{A}}=<M\otimes I>_{\rho_{AB}}\label{eq:Expt-1}
\end{equation}

For general two qubit systems, the Stokes vector is a $16$ parameter
real valued column vector, $S=[S_{00},S_{x0},S_{y0},S_{z0},S_{0x},S_{0y},S_{0z},\dotsb S_{zz}]^{T}$
where the entries in the column vector are the expectation values
of the generalized two qubit Pauli matrices, 
\begin{equation}
S_{i_{1}i_{2}}=\frac{1}{4}Tr(\rho\sigma_{i_{1}}\otimes\sigma_{i_{2}})\label{eq:Stokes_two-1}
\end{equation}

where $i_{1},\,i_{2}\in[0,\,x,\,y,\,z]$. Replace the operator $M$
in the Eq (\ref{eq:Expt-1}) with the Pauli's operators $\sigma_{i}$,
\[
<\sigma_{i}>_{\rho_{A}}=<\sigma_{i}\otimes I>_{\rho_{AB}}
\]

These expectation values are essentially the Stokes vector of the
first subsystem $S^{A}$, given by
\begin{equation}
S_{i}^{A}=Tr[\sigma_{i}\rho_{A}]=Tr[(\sigma_{i}\otimes I)\rho_{AB}]=S_{i0}\label{eq:Twotosingle-1}
\end{equation}

From Eq (\ref{eq:Twotosingle-1}) it is clear that the Stokes vector
of the first subsystem $S^{A}$ is part of the two qubit Stokes vector
$S^{AB}$, that is, $S_{i}^{A}=S_{i0}^{AB}$. The Stokes vector of
the subsystem can be obtained from that of the Stokes vector of the
global system by the construction of the transformation matrix $T_{1}$,
such that
\begin{equation}
S^{A}=T{}_{1}S\label{eq:TS-1}
\end{equation}

where $T{}_{1}$ is the $4\times16$ matrix given by $T{}_{1}=2[I\,O\,O\,O]$.

Let $W$ be the DWF of the the system $\rho_{AB}$. From Eq (\ref{eq:SHW}),
two qubit Stokes vector $S$ can calculated from $W$ by,

\begin{equation}
S=H_{2}W\label{eq:SHW-2}
\end{equation}

where $H_{2}$ is the $4^{2}\times4^{2}$ dimensional Hadamard matrix,
which is an element of the set $\mathcal{\mathbb{S}}_{2}^{H}$, that
is the set of Hadamard matrix for the two qubit systems. Based on
the quantum net of $W$, we can choose the Hadamard matrix from the
set $\mathcal{\mathbb{S}}_{2}^{H}$. Therefore, from Eq (\ref{eq:TS-1})
and Eq (\ref{eq:SHW-2}) one can calculate the Stokes vector of the
first subsystem directly from the DWF of the two qubit system by,
\begin{equation}
S^{A}=T{}_{1}H_{2}W\label{eq:S_one-1}
\end{equation}

where the product $T{}_{1}H_{2}$ is the $4\times16$matrix. The inverse
transformation from the Eq (\ref{eq:WHS}) takes the Stokes vector
to the DWF of the subsystem $A$, by
\begin{equation}
W^{A}=H_{1}^{-1}S^{A}\label{eq:WHS-2}
\end{equation}

where $H_{1}$ is the $4\times4$ Hadamard matrix, contained in the
set $\mathcal{\mathbb{S}}_{1}^{H}$ of single qubit system. Therefore
from the Eq (\ref{eq:S_one-1}) and Eq (\ref{eq:WHS-2}), the DWF
of the first subsystem can be given by,

\[
W^{A}=H_{1}^{-1}T{}_{1}H_{2}W^{AB}
\]

that is
\begin{equation}
W^{A}=P_{1}W^{AB}\label{eq:PT1-1}
\end{equation}

where $P_{1}=H_{1}^{-1}T{}_{1}H_{2}$ is a $4\times16$ matrix. Using
this relation one can calculate the DWF of the first subsystem from
the DWF of the two qubit system. That is Eq (\ref{eq:PT1-1}) performs
the reduction operation for the two qubit DWF. By the similar transformation
one can construct the reduction operation for the second subsystem
by suitable construction of the matrix $T{}_{2}$ as,
\begin{equation}
W^{B}=P_{2}W^{AB}\label{eq:PT2-1}
\end{equation}
where $P_{2}=H_{1}^{-1}T_{2}H_{2}$. Hence, the reduction operation
for the general two qubit DWFs can be performed using Eq (\ref{eq:PT1-1})
and Eq (\ref{eq:PT2-1}). Here, it is important to note that, the
reduction formula is general enough to compute the DWF of the subsystem
defined in any arbitrary quantum net from the DWF of the two qubit
system defined in an arbitrary quantum net. In our transformation
equations, the information about the quantum net of the global system
and that of the subsystems are contained in the Hadamard matrices
$H_{2}$ and $H_{1}$. So equations (\ref{eq:PT1-1}) and (\ref{eq:PT2-1})
carries out the reduction operation for chosen quantum net. The reduction
formula given by Holmes et al., in Eq (\ref{eq:WA1}) and Eq (\ref{eq:WBM})
are the special cases of this formula. As an aside we note that the
reduction formula can be used to calculate quantities of interest
like concurrence of a bipartite pure state defined by $W^{AB}$ as,
\[
C(W^{AB})=\sqrt{2\left(1-\underset{\alpha}{\sum}W_{\alpha}^{A^{2}}\right)}
\]

where $W_{\alpha}^{A}$ is the DWF of subsystem $A$.

\section{Reduction formula for the general multiqubit DWF}

In the density matrix formalism, from the given $n$-qubit state $\rho$,
the state of an arbitrary $k$-qubit subsystem can be calculated by
the partial trace operation. In this section, we derive a method of
performing the equivalent of a partial trace operation on the general
$n$-qubit DWF by ``tracing out'' $n-k$-qubits . In the DWF setting
this can be done using the following facts: the state of the sub-system
can be readily extracted from the Stokes vector of the composite state
and the transformation formula between Stokes vector and the DWF given
in Eq (\ref{eq:SHW}) can be used the obtain the DWF of the sub-system.
To see how this may be accomplished, consider a density matrix of
the general $n$-qubit system $\rho$ and let $M^{i}$ be some observable
acting on the $i$-th qubit. Given $\rho$, the expectation value
of the observable $M^{i}$ on $i$-th qubit can be calculated as:

\begin{align}
<M^{i}>_{\rho_{i}} & =Tr(M^{i}\rho_{i})\\
 & =Tr[(I\otimes I\otimes...\otimes M^{i}\otimes...\otimes I)\rho]\label{eq:ave_M-1}
\end{align}

This can be generalized for any $k$-partite subsystem. When this
problem is cast in terms of Stokes vectors, the observables are the
Pauli operators and the expectation values $<\sigma_{j}>_{\rho_{i}}$'s
are the Stokes parameters $S_{j}^{i}$ of the subsystem $i$. That
is, $S_{j}^{i}=<\sigma_{j}>_{\rho_{i}}$. But from Eq (\ref{eq:ave_M-1}),
it is clear that, $<\sigma_{j}>_{\rho_{i}}=<I\otimes I\otimes...\otimes\sigma_{j}^{i}\otimes...\otimes I>_{\rho}$.
This implies, 
\begin{equation}
S_{j}^{i}=S_{0...j...0}\label{eq:stoke1toN-1}
\end{equation}

Therefore, in the Stokes vector representation, the state of the $i$-th
subsystem is a part of the multiqubit Stokes vector $S$. Similarly,
for any $k$-partite subsystem, its state is contained in the multiqubit
Stokes vector from which it can be easily extracted. Consider the
case of tracing out the last $(n-1)$ subsystems from the $n$-qubit
state, giving the state of the first system $S^{1}$. The Stokes vector
of the first system can be calculated as,
\begin{equation}
S^{1}=T_{1}S\label{eq:S1TS-1}
\end{equation}

where $T_{1}$ is the $4\times4^{n}$ matrix, given by $T_{1}=2^{n-1}[I\,O\,O\,...\,O]$
with $I$ is a $4\times4$ identity matrix and $O$ is a $4\times4$
matrix with all entries being zero. Similarly any $k$-qubit Stokes
vector can be constructed from the Stokes vector of the multiqubit
systems with the help of a suitable transformation matrix $T_{k}$.
\begin{equation}
S^{k}=T_{k}S\label{eq:SkTS-1}
\end{equation}

where $T_{k}$ is a $4^{k}\times4^{n}$ matrix. For a given multiqubit
DWF $W$, the corresponding Stokes vector can be calculated using
the $4^{n}\times4^{n}$ Hadamard matrix $H_{n}\in\mathcal{\mathbb{S}}_{n}^{H}$
by,
\begin{equation}
S=H_{n}W\label{eq:SHW-1-1}
\end{equation}

where the Hadamard matrix $H_{n}\in\mathcal{S}_{n}^{H}$. Let $\mathcal{\mathbb{S}}_{n}^{H}$
be the set containing $N^{N+1}$ possible Hadamard matrices associated
with each quantum net. Therefore, from Eqs (\ref{eq:SkTS-1}) and
(\ref{eq:SHW-1-1}), 
\begin{equation}
S^{k}=T_{k}H_{n}W\label{eq:S1THW-1}
\end{equation}

where the subscript $n$ of $H_{n}$ indicate that the Hadamard matrix
picked up from the set $\mathcal{\mathbb{S}}_{n}^{H}$ of the $n$-qubit
systems. Eq (\ref{eq:S1TS-1}) allows us to calculate the Stokes vector
of the $k$-qubit subsystem from the multiqubit DWF. Here the knowledge
of the quantum net of $W$ is implicitly available in the Hadamard
matrix $H_{n}$. Using the inverse formula given in Eq (\ref{eq:WHS}),
we can find the DWF of the $k$-qubit system as,
\begin{equation}
W^{k}=H_{k}^{-1}T_{k}H_{n}W
\end{equation}

where $H_{k}\in\mathcal{\mathbb{S}}_{k}^{H}$.

\begin{equation}
W^{k}=P_{k}W\label{eq:WptN-1}
\end{equation}

Therefore, Eq (\ref{eq:WptN-1}) helps us perform the reduction operation
for the multiqubit DWF. Thus, we find that the quantum net of the
global system and the subsystems are to be obtained from the Hadamard
matrices $H_{n}\in\mathcal{\mathbb{S}}_{n}^{H}$ and $H_{k}\in\mathcal{\mathbb{S}}_{n}^{H}$
respectively. Hence, the choice of the quantum net of the global system
and that of the reduced system can be freely made by an appropriate
choice of the corresponding $H_{n}$ and $H_{k}$.

$\vphantom{}$

\section{Conclusions}

There are many contexts in the fields of quantum computation and quantum
information where access to subsystem information is vital. The quantification
of entanglement present in a composite bipartite system through Concurrence
and the derivation monogamy relationships from tripartite entangled
states are typical examples. Similarly, in the case of multiqubit
systems, the distribution of entanglement over suitably partitioned
subsystems is a problem of interest. Frequently, one also requires
to enlarge the Hilbert space by taking a tensor product of the system
with that of the environment, subjecting the joint system to a unitary
evolution and eventually tracing out either the environment or the
system. The theory of POVMs and weak measurements are typical examples
of such procedures. Hitherto, such techniques have been uniquely applied
to the case where the state of the system is represented in terms
of the density matrix and an equivalent approach was not available
at least in the case of systems represented by the discrete Wigner
function. While the representation of the state of continuous systems
by Wigner functions has found widespread use, this not the case for
the DWF due to some obvious limitation. An important limitation with
DWF as stated earlier has been the absence of a general reduction
formula, which problem has been addressed in the present work. While
its true that DWF, density matrix and Stokes vector representations
are but linear transforms of each other, experimental situations could
make one choice or the other more favorable and experimental reconstructions
of the the different representation are also different. Going by the
experience with continuous system, where the phase space representation
of the state provides certain unique insights, further development
of its discrete analog is warranted. Motivated by such considerations,
the present work is a step in the direction of developing the relevant
tools for the Discrete Wigner function of multiqubit systems.
\begin{acknowledgments}
K. Srinivasan acknowledges Indira Gandhi centre for atomic research,
DAE for the award of research fellowship.
\end{acknowledgments}

\bibliographystyle{apsrev4-1}

%

\end{document}